\documentclass[fleqn]{2023SCGE}

\newcommand       \g          {\,{\rm g}}

\newcommand       \mum           {\,{\rm \mu m}}

\newcommand       \s            {\,{\rm s}}

\newcommand       \J            {{\rm J}}

\newcommand       \K            {{\rm K}}

\newcommand   \apjs       {ApJs}

\usepackage{graphicx} 
\usepackage{epstopdf}
\usepackage{threeparttable}
\usepackage{rotating}
\usepackage{adjustbox}
\usepackage{hyperref}

\begin{document}

\ensubject{subject}

\ArticleType{Article}
\Year{2024}
\Month{June}
\Vol{66}
\No{1}
\DOI{??}
\ArtNo{000000}
\ReceiveDate{June 11, 2024}

\title{Temporal-spatial distribution of YSOs nearby Taurus region}

\author[1,2]{Jiaming Liu}{{jmliu@hebtu.edu.cn}}%
\author[3]{Min Fang}{mfang@pmo.ac.cn}
\author[4]{Chao Liu}{}
\author[1,2]{Xiaolong  Wang}{}%
\author[1,2]{Wenyuan Cui}{}
\author[5]{Licai Deng}{}

\AuthorMark{Liu J M}

\AuthorCitation{Liu J M, Fang M, Liu C, et al}

\address[1]{Department of Physics and Hebei Advanced Thin Films Laboratory, Hebei normal University, Shijiazhuang 050024, China}
\address[2]{Shijiazhuang Key Laboratory of Astronomy and Space Science, Hebei normal University, Shijiazhuang 050024, China}
\address[3]{Purple Mountain Observatory, Chinese Academy of Sciences, Nanjing 210023, China}
\address[4]{Key Laboratory of Space Astronomy and Technology, National Astronomical Observatories, CAS, Beijing, 100101, China}
\address[5]{Key Laboratory of Optical Astronomy, National Astronomical Observatories, CAS, Beijing 100101, China}

\abstract{The Taurus region is one of the most extensively studied star-forming regions. Surveys indicate that the young stars in this region are comprised of Young Stellar Objects (YSOs) that cluster in groups associated with the molecular cloud (Grouped Young Stellar Objects, GYSOs), and some older ones that are sparsely distributed throughout the region (Distributed Young Stellar Objects, DYSOs). To bridge the age gap between the GYSOs ($\le$5\,Myr) and the DYSOs (10-20\,Myr), we conducted a survey to search for new YSOs in this direction. Based on infrared color excesses and Li I absorption lines, we identified 145 new YSOs. Combining these with the previously identified GYSOs and DYSOs, we constructed a sample of 519 YSOs that encompasses the entire region. Subsequently, we calculated the ages of the sample based on their proximity to the local bubble. The age versus Distance to the Local Bubble ($D_{\rm LB}$) relationship for the DYSOs shows a clear trend: the farther they are from the Local Bubble, the younger they are, which is consistent with the supernovae-driven formation scenario of the Local Bubble. The GYSOs also exhibit a mild age versus $D_{\rm LB}$ trend. However, they are significantly younger and are mostly confined to distances of 120 to 220\,pc. Considering their distribution in the age versus $D_{\rm LB}$ space is well separated from the older DYSOs, they may also be products of the Local Bubble but formed in more recent and localized events.}


\maketitle

\begin{multicols}{2}
\section{Introduction}\label{section1}
As part of the Local Bubble, centered at the Sun with a radius of roughly 300 pc \cite{ref1}, the Taurus star-forming region (SFR) is located approximately 145 pc to the west of the Sun \cite{ref2}, and covers an extension of more than 100~deg$^{2}$ on the sky \cite{ref3}. As a typical SFR of low stellar densities, its proximity to the Sun favors observations and makes Taurus one of the most studied nearby star-forming regions. Surveys for young stellar objects (YSOs) in the Taurus region have been performed for over 30 years and have nearly covered all bands of observations, e.g., X-ray, UV (ultraviolet), optical, and infrared \cite{ref4,ref5,ref6,ref7,ref8,ref9,ref10,ref11,ref12}. Based on the astrometric data from Gaia \cite{ref13,ref14,ref15}, Luhman (2018) \cite{ref16} and Esplin \& Luhman (2019) \cite{ref3} studied the clustering properties of the identified YSOs in Taurus. They found that the YSOs in Taurus are kinematically correlated and that most of them concentrate in several dense sites within the Taurus molecular cloud. Using these grouping properties, they re-examined the YSO candidates previously identified in the literature. They confirmed the identities of 532 YSOs \cite{ref17} based on the criteria that the new YSOs should be spatially close to and kinematically associated with the previously known YSOs in Taurus. These stars, which are young (1-5 Myr), are generally categorized into 13 groups based on their distances and ``proper motion offsets'' \footnote{The ``proper motion offsets'' of the YSOs refer to the difference between the observed proper motion and the expected proper motion, which is derived from the celestial coordinates and distance of the star, assuming a specified space velocity (for the GYSOs in the Taurus region, the space velocity is U, V, W = -16, -12 , -9 $\rm km\,s^{-1}$) \cite{ref18}} (hereafter referred as the GYSOs).

Surveys of the Taurus region have also found that there are a certain number of YSOs that are displaying clear evidence of youth but are sparsely distributed in the region (hereafter, DYSOs) \cite{ref4,ref6,ref19,ref20,ref21}. Kraus et al. (2017) \cite{ref18} studied these stars and found that they are relatively old (10-20~Myr) and disk-free (YSOs without circumstellar disks). They studied the proper motions of the DYSOs and found that they are broadly co-moving with the GYSOs of Taurus. Thus, they conclude that these stars might be the formerly formed populations of Taurus. Luhman (2018) \cite{ref16} also discussed the correlation between the DYSOs and the young GYSOs of Taurus, and calculated the 10\,Myr relative drifts of the DYSOs. The results show that the values are too large for most of them to be associated with the Taurus cloud. The ambiguous relation between the DYSOs and the GYSOs demonstrates that their formation might be different. Besides, the age gap between the DYSOs ($\sim$10-20\,Myr) and the GYSOs($\le$5\,Myr) is also worth noting, and it drives us to postulate that there may be some young stars of 5-10\,Myr that have not been discovered yet. These stars may offer us the opportunity to study the relationship between the GYSOs and the DYSOs, and to understand the star formation processes in the Taurus region. To address these issues, conducting a survey to identify new YSOs in the Taurus region could be beneficial.

We have structured our work as follows: In Section 2, we describe the data selection and the identification of new YSOs. In Section 3, we will discuss the temporal-spatial distribution of the YSOs in Taurus in relation to the Local Bubble, the Per Tau shell, and the Tau ring. This will be followed by a brief discussion in Section 4. Finally, we will provide a concise summary in Section 5.

\section{Data selection and analysis}\label{sec2}
\begin{figure}[H]
\centering
\vspace{-0.3cm}
\setlength{\abovecaptionskip}{-0.06cm}
\includegraphics[scale=0.32]{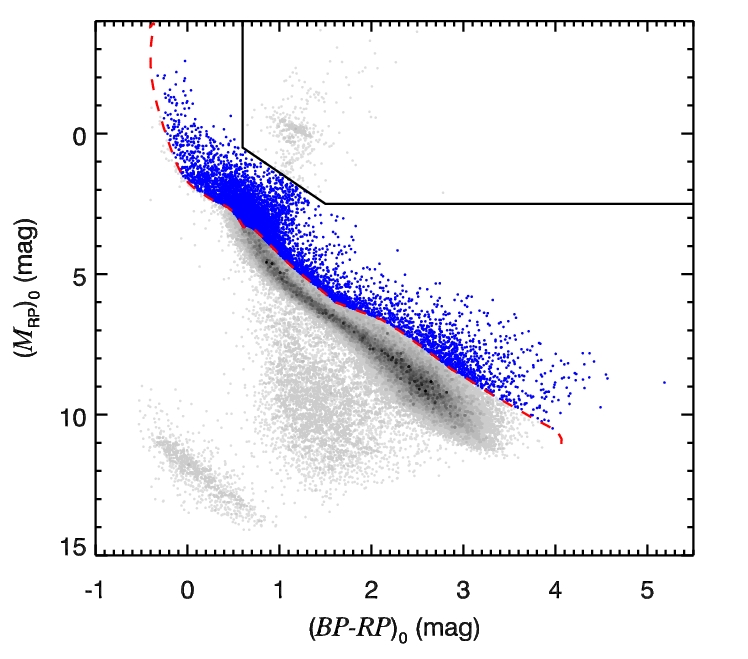}
\caption{The $G_{BP}-G_{RP}$ vs. $M_{G_{RP}}$ color-magnitude diagram shows the stars (gray dots) and includes the 20\,Myr isochrone to eliminate field stars. The red dashed line denotes the 20\,Myr isochrone of the PARSEC model \cite{ref23} with solar metallicity \cite{ref24}. The black solid line is defined by the points (0.6, -5.0), (0.6, 0.5), (1.5, 2.5), (5.6, 2.5), and (5.6, -5.0) to remove evolved stars. The selected candidates are shown as blue solid dots.}
\label{CMDS}
\end{figure}

The primary data for this work was taken from \emph{Gaia} DR3 \cite{ref14}. To cover the entire Taurus region, we set the search area to $55^{\circ} \leq \text{RA} \leq 90^{\circ}$ and $10^{\circ} \leq \text{DEC} \leq 35^{\circ}$. To ensure that the stars we are interested in are associated with the Taurus region, only stars within 100 to 300\,pc from the Sun were selected \cite{ref22}. We excluded stars in the region $55^{\circ} \leq \text{RA} \leq 70^{\circ}$ and $30^{\circ} \leq \text{DEC} \leq 35^{\circ}$ to remove stars from the Perseus region. For safety, only stars with signal-to-noise ratios in parallax greater than 5 ($\varpi/\sigma_{\varpi} \geq 5$) and photometric errors in $G_{BP}$ and $G_{RP}$ bands less than 5\% were retained. As we are interested in YSOs, an age cut of 20 Myr was applied in the CMD (color-magnitude diagram) to eliminate field stars (see Figure \ref{CMDS}). We considered the PARSEC model isochrones from Bressan et al. (2012) \cite{ref23,ref24} for this step. We selected only those stars that are above the 20\,Myr isochrone and below the black solid line, which is visually defined by the points (0.6, -5.0), (0.6, 0.5), (1.5, 2.5), (5.6, 2.5), and (5.6, -5.0) to exclude evolved stars. The extinctions of stars were corrected based on the 3D extinction map of Green et al. (2019) \cite{ref25} and the extinction law of Wang \& Chen (2019) \cite{ref26}. Based on these criteria, 7046 stars were selected, and their identities will be further evaluated.

\subsection{Photometry and spectra}
To estimate the extinctions of the stars, the photometric data in the $G_{\rm BP}$ and $G_{\rm RP}$ bands of {\em Gaia}, the optical bands $g_{\rm PS1}$, $r_{\rm PS1}$, $i_{\rm PS1}$, $z_{\rm PS1}$, $y_{\rm PS1}$ of Pan-STARRS (Panoramic Survey Telescope and Rapid Response System \cite{ref53}), the photometric data in the $g_{\rm SDSS}$, $r_{\rm SDSS}$, $i_{\rm SDSS}$, $z_{\rm SDSS}$ bands of SDSS9 \cite{ref54}, the $B,V,u,g,r,i,z$ bands of the APASS9 (AAVSO Photometric All-Sky Survey \cite{ref55}), the $B_{\rm T}$ and $V_{\rm T}$ bands of Tycho-2 \cite{ref56}, and the near-infrared photometric data $J$, $H$, and $K_{\rm s}$ of the Two-Micron All Sky Survey (2MASS \cite{ref27}) are considered. The near- and mid-infrared photometric data from $W1$ (3.4 $\mu$m), $W2$ (4.6 $\mu$m), and $W3$ (12 $\mu$m) of WISE (the Wide-field Infrared Survey Explorer \cite{ref28}) are employed to search for stars with disks (stars surrounded by circumstellar disks).

To estimate the extinctions of the stars, the $G_{\rm BP}$, $G_{\rm RP}$ bands of {\em Gaia}, the optical bands $g_{\rm PS1}$, $r_{\rm PS1}$, $i_{\rm PS1}$, $z_{\rm PS1}$, $y_{\rm PS1}$ of Pan-STARRS (Panoramic Survey Telescope and Rapid Response System \cite{ref53}), the $g_{\rm SDSS}$, $r_{\rm SDSS}$, $i_{\rm SDSS}$, $z_{\rm SDSS}$ bands of SDSS9 \cite{ref54}, the $B,V,u,g,r,i,z$ bands of the APASS9 (AAVSO Photometric All-Sky Survey \cite{ref55}), the $B_{\rm T}$ and $V_{\rm T}$ bands of Tycho-2 \cite{ref56} and the near-infrared bands $J$, $H$ and $K_{\rm s}$ of the Two-Micron All Sky Survey (2MASS \cite{ref27}) are considered. The near- and mid-infrared photometric data of the $W1$ (3.4 $\mum$), $W2$ (4.6 $\mum$) and $W3$ (12 $\mum$) from WISE (the Wide-field Infrared Survey Explorer \cite{ref28}) are employed to search for stars with circumstellar disks.

Since the Li~I absorption line is a key diagnostic tool for identifying young late-K and M type stars, we have utilized the low-resolution spectral data from LAMOST~\cite{ref29} in its seventh data release (DR7). These spectra span a broad wavelength range from approximately 3900 to 9000~\AA, boasting a resolution of roughly 1800.

\begin{figure}[H]
\centering
\hspace{-0.8cm}
\setlength{\abovecaptionskip}{0.16cm}
\includegraphics[scale=0.32]{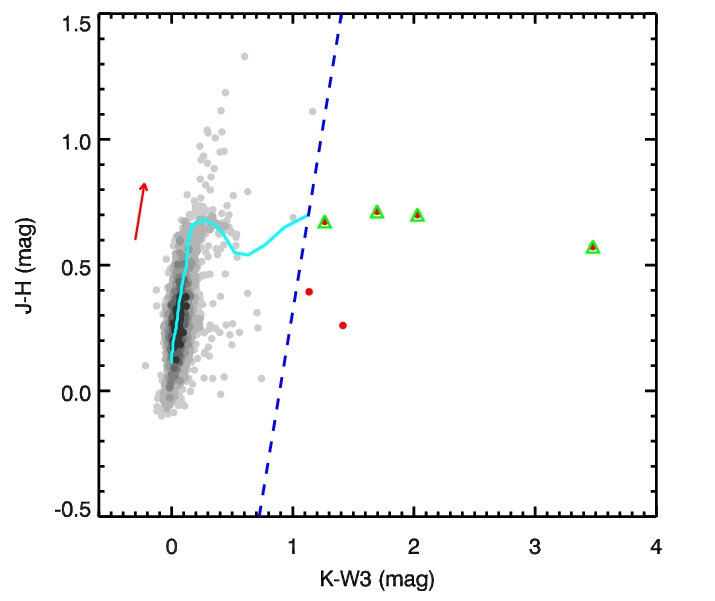}
\caption{Color-color diagrams for the YSO candidates (gray dots). The intrinsic colors of pre-main-sequence (PMS) stars are derived from Pecaut \& Mamajek (2013) \cite{ref45} and are indicated by a cyan solid line. The red arrow represents the extinction vector for $A_{V}=2.0$ mag. A blue dashed line, which passes through the end of the intrinsic PMS colors, is parallel to the extinction vector. Consequently, the color excesses of stars (red solid dots) located to the right of this line cannot be accounted for by interstellar extinction alone. The newly identified disk-bearing stars are marked with green triangles.}
\label{CCD}
\end{figure}

\subsection{Identification of new YSOs}\label{secidt}
As mentioned above, we identified a sample of 7,046 candidate young stars within the Taurus region. Among them, 388 stars have previously been identified \cite{ref16,ref3,ref2,ref37,ref18}. The identities of the remaining 6,658 stars will be further evaluated.

\begin{figure}[H]
\centering
\vspace{-0.5cm}
\setlength{\abovecaptionskip}{0.1cm}
\includegraphics[scale=0.35]{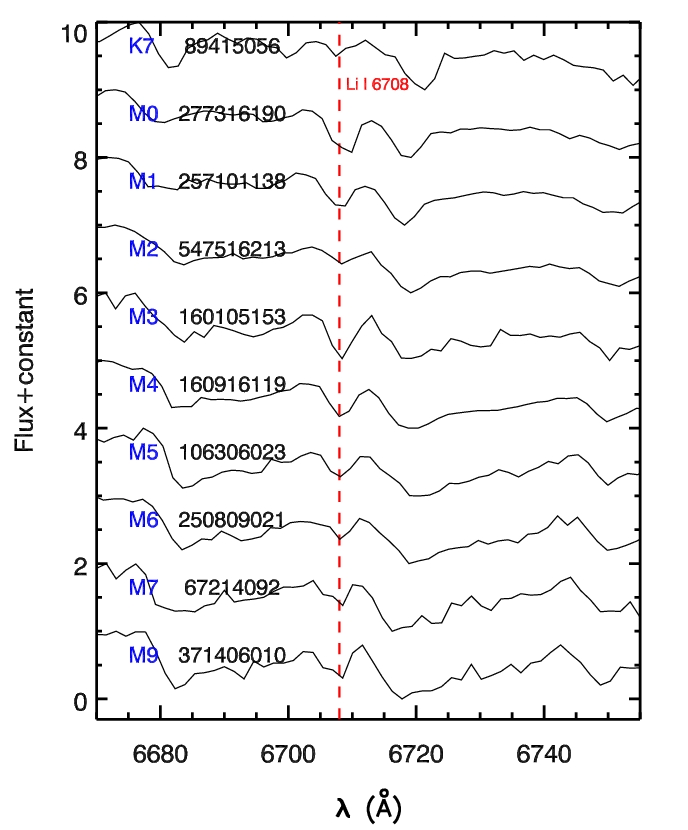}
\caption{The spectra from LAMOST of the newly identified YSOs are shown, along with their corresponding ``obsid'' from LAMOST. The red dashed line indicates the Li\,I absorption line at 6708\AA.}
\label{Li}
\end{figure}

\subsubsection{Searching for Disk bearing YSOs}
As protostars continue to contract, accretion disks form around them due to the conservation of angular momentum. Because of the emission from the disk, young stellar objects (YSOs) typically exhibit excess emission in the near- and mid-infrared bands. This property is often used to characterize YSOs. In this work, we utilized the photometric data from the near- and mid-infrared bands of the 2MASS and ALLWISE Source Catalogues to search for disk-bearing stars. Among the 6,658 YSO candidates, we found 6,601 common sources in both catalogues with a cross-matching radius of 2$''$. After excluding stars with poor photometric quality, which are those with ``ph\_qual'' (photometric quality flag) worse than ``B'' in the 2MASS and WISE bands, and those with photometric uncertainties greater than 0.05 mag in the $J$, $H$, $K_{\rm S}$, $W1$, $W2$, and $W3$ bands, 4,062 stars remained.

We analyzed the colors of these stars and identified 6 stars that showed color excess in the mid-infrared bands (see Figure \ref{CCD} for details). We then reviewed the Spectral Energy Distributions (SEDs) and WISE images for each source. Two stars were found to be in close proximity to infrared bright stars and were consequently excluded for safety reasons. The remaining four stars, which exhibit clear excess emission (greater than 3-sigma) in the infrared, are newly identified in this study (see Table 1 for details).

\subsubsection{YSOs with Lithium line}
The Li I absorption line at 6708\,\AA\ is also a strong indicator of youth for late K- and M type stars \cite{ref35}, particularly for those with spectral types later than K7. Among the 6658 YSO candidates, 5201 have been released with low-resolution spectra in LAMOST DR7. We examined the spectra of candidates with spectral types later than K7 and found that 141 of them exhibit a clear Li I absorption line (with equivalent widths greater than 50\,$\rm m\AA$; see Figure \ref{Li} for an example).

\begin{figure}[H]
\centering
\hspace*{-0.6cm}
\vspace{-0.5cm}
\setlength{\abovecaptionskip}{0.5cm}
\includegraphics[scale=0.36]{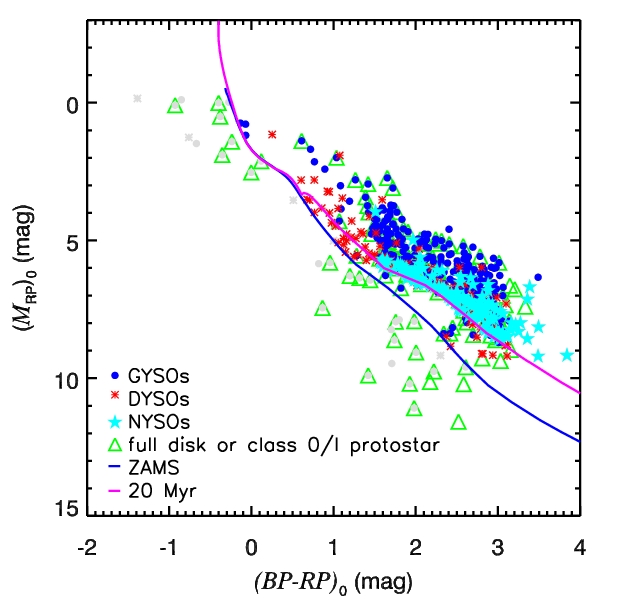}
\caption{The dereddened color-magnitude diagram for the NYSOs in this work is shown as cyan five-pointed stars. Also plotted are the DYSOs (red asterisks) and GYSOs (blue solid dots) with high photometric quality (i.e., stars with $e\_F_{BP}/F_{BP} \le 0.1$ and $e\_F_{RP}/F_{RP} \le 0.1$) from the literature. The class 0/I or full-disk YSOs classified by Esplin, Luhman \& Mamajek (2014) \cite{ref11} are marked as green triangles. The blue solid line and purple solid line represent the zero age main sequence (ZAMS) of solar metallicity \cite{ref63} and the 20\,Myr isochrone from the PARSEC model \cite{ref23}, respectively.}
\label{cds}
\end{figure}

\subsection{Color Magnitude Diagram of the YSOs in Taurus}

By examining the infrared excess and the Li I lines in the spectra of the candidate stars, we have identified 145 new YSOs (hereafter referred to as NYSOs), and their general characteristics are presented in Table \ref{tbl1}.

In Figure \ref{cds}, we present the dereddened CMD of these NYSOs as cyan five-pointed stars. To enhance the extinction estimation for them, we compiled and cross-matched archival data from Pan-STARRS, SDSS, APASS9, Tycho-2, and 2MASS within a 2'' radius. After removing saturated bands, we constructed a catalog with at least two visual bands and $J$-band photometry for all 145 NYSOs in this work. The extinction values for each star were then determined by minimizing the $\chi^{2}$ difference between the observed SED and model SEDs \cite{ref57}, as detailed in Table \ref{tbl1}. The identified YSOs from the literature are also plotted for comparison, with GYSOs \cite{ref17} and DYSOs (\cite{ref1,ref5,ref6,ref7,ref9,ref10,ref11,ref18,ref19,ref20,ref21}; see Krolikowski et al. 2021 \cite{ref37} and references therein) marked as blue solid dots and red asterisks, respectively. To eliminate the impact of measurement quality, we performed photometric screening on the previously identified YSOs, selecting only stars with reported parallactic distances \cite{ref22}, $e\_F_{BP}/F_{BP} \le 0.1$, and $e\_F_{RP}/F_{RP} \le 0.1$, and we also recalculated their extinction with SED fitting. The figure shows that the NYSOs are mostly late-type stars (K- and M-type) and comparatively old, which is as expected.

\subsection{Relationship between the new identified YSOs and the previous identified YSOs of Taurus}
\begin{figure}[H]
\centering
\hspace*{0.1cm}
\vspace{-0.35cm}
\includegraphics[scale=0.23]{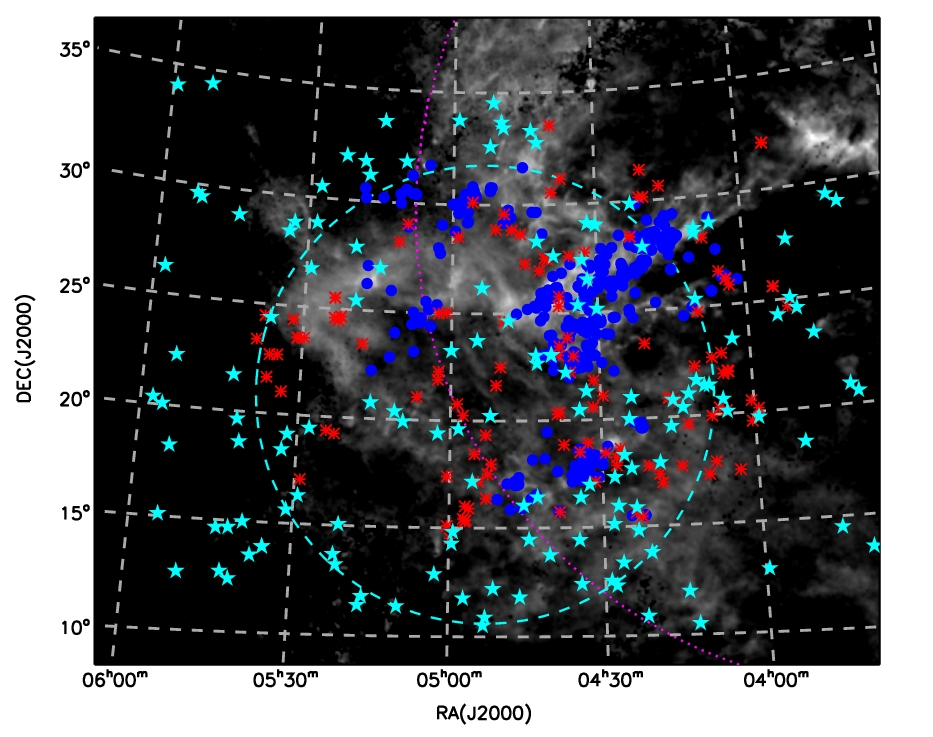}
\caption{The locations of the YSOs identified in this work are shown as cyan five-pointed stars. Also plotted are the identified GYSOs \cite{ref17} (blue solid dots) and DYSOs (red asterisks). The peripheries of the Per Tau shell and Tau ring are marked with a purple dotted line and a cyan dashed line, respectively. The background contour represents the 0-300\,pc cumulative extinction from Green et al. (2019) \cite{ref25}.}
\label{loc}
\end{figure}

In Figure \ref{loc}, we show the spatial distribution of the NYSOs. For comparison, the identified GYSOs and DYSOs with parallax measurements from {\em Gaia} DR3 are also displayed. The figure reveals that the majority of YSOs are clustered in several regions of dense interstellar medium. These stars are predominantly GYSOs, which are younger and share a common motion \cite{ref17}. In contrast, the NYSOs and DYSOs are sparsely distributed in areas of low extinction.
Luhman (2023) \cite{ref17} explored the clustering characteristics of YSOs within the Taurus molecular cloud complex. Utilizing proper motion and parallax data, they computed the ``proper motion offsets'' of the YSOs, which are regarded as an effective way to mitigate the projection effect. Based on the distances and proper motion offsets of the stars, they identified that the GYSOs in the Taurus region can be separated into 13 distinct ``kinematic clusters'', whereas most DYSOs are too distant to be considered part of these clusters.

\begin{table}[H]
\footnotesize
\centering
\caption{New YSOs identified in this work.}
\label{tbl1}
\begin{threeparttable}
\begin{tabular}{lcccccccccc}
\hline\hline
GAIA & obsid$^{\rm a}$ &  R.A. J2000 &  Decl. J2000 & SpT & $A_V$ & method & $EWs_{Li}$ & $J-K_{\rm S}$ & $K-W3$&G/D$^{\rm b}$ \\
\hline
&&deg&deg&&mag&&$\rm \AA$&mag&mag&\\
\hline
   3421038487461634432 &  &        78.744861 &        26.997857 &  &     0.00 & disk/L1517 &  &      0.94 &       1.26 & G\\
   3306698349143972096 &  &        68.618822 &        12.396331 &  &     0.01 & disk &  &      1.00 &    1.70 & D\\
   3443849299289015424 &  &        88.033247 &        29.487062 & M5 &      0.01 & disk &   &      0.97 &       2.03 & D\\
   3304172564776041728 &  &        63.578048 &        11.970208 & G6 &    0.00 & disk &  &      0.92 &       3.48 & D\\
    181939938557599104 &     89415056 &        78.715996 &        33.622385 & K7 &       0.00 & Li &    -0.09 &      0.77 &     0.10 & D\\
     51884824140206720 &    277316190 &        61.303229 &        20.130234 & M0 &      0.02 & Li &     -0.42 &      0.84 &      0.17 & D\\
   3340935388886477696 &    257101138 &        82.329144 &        12.158156 & M1 &     0.00 & Li &     -0.28 &      0.86 &     0.02 & D\\
   3403412182197974400 &    547516213 &        84.653470 &        22.113817 & M2 &     0.01 & Li &     -0.23 &      0.90 &      0.32 & D\\
     53860268579929984 &    160105153 &        60.838487 &        23.428837 & M3 &     0.01 & Li &     -0.68 &      0.93 &      0.33 & D\\
   3389329980866394112 &    160916119 &        80.370807 &        13.210594 & M4 &     0.00 & Li &     -0.51 &      0.89 &      0.39 & D\\
   3417628249070016256 &    106306023 &        79.920225 &        25.445293 & M5 &      0.02 & Li &     -0.45 &      0.89 &      0.64 & D\\
   3408524670748967808 &    250809021 &        74.594470 &        19.613998 & M6 &     0.01 & Li &     -0.38 &      0.89 &      0.49 & D\\
    161422914142980096 &     67214092 &        70.875427 &        33.324589 & M7 &      0.04 & Li &     -0.47 &      0.88 &      0.75 & D\\
   3313784735944668544 &    371406010 &        66.912212 &        17.308816 & M9 &      0.02 & Li &     -0.73 &      0.84 &      0.76 & D\\
   3421550069606804224 &     15505226 &        80.011872 &        27.902244 & M4 &       0.00 & Li &     -0.13 &           --&      -- & D\\
     52262849980581376 &    249411234 &        62.348995 &        21.375181 & M9 &      0.05 & Li &    -0.07 &      0.96 &      0.67 & D\\
   3445519246998786176 &     14505069 &        82.092518 &        28.890364 & M7 &     0.00 & Li &     -0.28 &      0.85 &      0.46 & D\\
    180313176745613056 &    173708042 &        79.454365 &        31.188547 & M6 &      0.03 & Li &     -0.72 &      0.90 &      0.28 & D\\
   3341538436655619840 &    404510207 &        84.374963 &        13.492860 & M6 &      0.02 & Li &     -0.58 &      0.86 &      0.48 & D\\
   3429888834512835456 &      8507200 &        89.492815 &        26.243084 & M6 &      0.01 & Li &     -0.38 &      0.91 &      0.79 & D\\
\hline\hline
\end{tabular}
\begin{tablenotes}
\footnotesize
\item[a] The ``obsid'' of LAMOST.
\item[b] ``G'' marks GYSOs while ``D'' for DYSOs.
\end{tablenotes}
This table is available in its entirety in machine-readable form.
\end{threeparttable}
\end{table}
\clearpage

\begin{figure}[H]
\vspace{-0.8cm}
\setlength{\abovecaptionskip}{-0.06cm}
\includegraphics[scale=0.36]{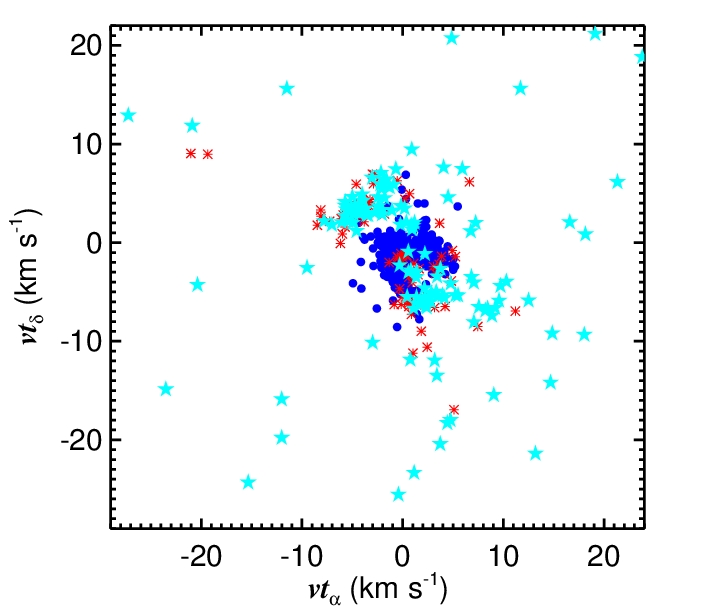}
\caption{The tangential velocities of the YSOs identified in this work are shown as cyan solid five-pointed stars. Also plotted are the identified GYSOs in Taurus (blue solid dots) and the DYSOs (red asterisks).}
\label{CPM}
\end{figure}

To correct for the projection effect, we first convert the proper motion of the star relative to the Sun into proper motion relative to the Local Standard of Rest (i.e., the solar motion \cite{ref64} corrected proper motion of star), and then we convert the proper motion into tangential velocity, which is displayed in Figure \ref{CPM}.
As can be seen from the figure, the GYSOs are clustered around (1.5, -2)\,$\rm km\,s^{-1}$, while DYSOs are diffusely distributed. Using this method, we examined the kinematics of the NYSOs. As depicted in Figure \ref{CPM}, the NYSOs (cyan five-pointed stars) are far from those of the GYSOs and are sparsely distributed. Additionally, we analyzed the spatial separation between the NYSOs and the groups delineated by Luhman (2023) and discovered that almost none of the NYSOs (except for one star, $Gaia$ 3421038487461634432, which coincides with members of group ``L1517'' both in position and kinematics) are positioned within 3 parsecs of the cluster members, suggesting that the NYSOs may not be associated with the kinematic clusters of Luhman (2023). In contrast, the NYSOs overlap spatially with the DYSOs (as shown in Figure \ref{loc}) and exhibit similar kinematics (as shown in Figure \ref{CPM}), indicating that they are more likely to be kinematically associated with the DYSOs of Taurus.

\section{Temporal-spatial distribution of the YSOs}\label{TSD}

In Section \ref{sec2}, we have identified 145 new YSOs. Through the analysis of their positions and kinematics, we discovered that nearly all of the newly identified YSOs are DYSOs. This finding is particularly intriguing when contrasted with GYSOs, as the distinctions between DYSOs and GYSOs in terms of position, kinematics, and age suggest that their formation processes are likely independent of each other.

Recently, Zucker et al. (2022) \cite{ref50} conducted an exhaustive investigation of the Local Bubble. They noted that almost all proximate star-forming regions (SFRs), including Taurus, Lupus, and Perseus, etc., are located at the edge of this bubble. Moreover, the kinematics of young stars within these SFRs exhibit a pronounced trend of radial motion, which is perpendicular to the surface of the Local Bubble. This pattern implies a possible connection between the processes of star formation in the solar neighborhood and the dynamic behavior of the Local Bubble. As a result, conducting an analysis to examine the temporal-spatial distribution of the YSOs in the context of the nearby bubble structures, i.e., the Local Bubble, the Per~Tau shell and the Tau ring \cite{ref62}, would be beneficial for enhancing our understanding of the star formation processes in the Taurus direction.


\subsection{Relationship between the Local bubble and the YSOs}

The Local Bubble is a structure centered on the Sun with a radius of $\sim$300 pc \cite{ref1}. Observations indicate that the interior of the bubble is filled with a hot ($\sim10^{6}$ K) and low-density interstellar medium \cite{ref49}, suggesting that the Local Bubble may have been formed by a series of supernova explosions \cite{ref50}. This supernova-driven formation theory of the Local Bubble is corroborated by observational studies. Based on the Hipparcos catalogue, Bouy \& Alves (2015) \cite{ref52} constructed a 3D map of the spatial density of nearby OB stars and noticed three large-scale stream-like structures: the Scorpius-Canis Majoris, Vela, and Orion streams. They found that the further an association is from the Local Bubble, the younger it tends to be. Most strikingly, the associations of the Orion stream are located about 100-400 pc from the Sun and span an age range of $\sim$2-20 Myr, exhibiting a nearly linear trend of age versus $D_{\rm LB}$ (distance to the Local Bubble).

The Taurus region, as a typical star-forming region near the Sun, is also part of the Local Bubble. Investigating whether star formation in this direction is related to the formation of the Local Bubble is of great interest. If star formation is indeed driven by the expansion of the Local Bubble, we would expect to observe a sequential formation trend among the young stars in that direction. To address this issue, it is necessary to accurately estimate the ages of the stars, which requires a high-quality sample. Therefore, we have conducted a certain level of screening on the identified YSOs, including GYSOs and DYSOs identified in the literature, as well as the newly identified YSOs from this work. We then excluded samples with spectral types later than M6 to eliminate brown dwarfs. It is known that an edge-on disk or variability can cause a star to appear in an abnormal location on the CMD (see the gray dots in Figure \ref{cds}). To account for this, we considered the zero age main sequence (ZAMS) of solar metallicity \cite{ref63} (indicated by the blue solid line in Figure \ref{cds}) and removed stars below the ZAMS for safety. Thus, in total, we constructed a sample of 519 YSOs (refer to Table \ref{tbl2} for more details).

\begin{table}[H]
\tiny
\centering
\caption{Compiled YSOs of Taurus direction.}
\label{tbl2}
\begin{threeparttable}
\begin{tabular}{lcccccc}
\hline\hline
GAIA & R.A. J2000 &  Decl. J2000 & SpT & $A_V$ & ref.$^{\rm a}$ & G/D$^{\rm b}$ \\
\hline
&deg&deg&&mag&&\\
\hline
     63408010940506496 &        55.045233 &        20.574855 & K7 &    0.00 & 5 & D\\
     41108064000723968 &        55.097387 &        13.578900 & K7 &    0.00 & 5 & D\\
    119945521612772992 &        55.161717 &        29.090041 & K7 &       0.00 & 5 & D\\
     63455053718627584 &        55.390731 &        20.924116 & K7 &       0.00 & 5 & D\\
    120051895065656576 &        55.671701 &        29.461620 & M1 &    0.00 & 5 & D\\
     39782911970360192 &        56.402148 &        14.528071 & K7 &     0.00 & 5 & D\\
     66755852048103552 &        57.570641 &        24.573383 & K7 &    0.00 & 5 & D\\
     50679072202242816 &        57.720017 &        18.493732 & M0 &     0.00 & 5 & D\\
     70807281817394048 &        57.852350 &        27.702710 & K7 &    0.00 & 5 & D\\
     66866735223542912 &        57.864339 &        25.061774 & M1 &     0.00 & 5 & D\\
     66665696393227904 &        58.009322 &        24.663245 & K5 &       0.00 & 4 & D\\
     66584538690731904 &        58.528160 &        24.333786 & M0 &     0.00 & 5 & D\\
    168834825106833152 &        58.625716 &        32.051203 & G5 &       0.00 & 4 & D\\
     67284446561797376 &        58.648167 &        25.619771 & K3 &       0.29 & 4 & D\\
     51369428065705600 &        59.791078 &        20.160036 & M4.75 &        1.89 & 3 & D\\
     50503906255772288 &        59.816210 &        19.763612 & M4 &       0.00 & 5 & D\\
     38425633585623808 &        59.856745 &        12.845081 & M4 &     0.00 & 5 & D\\
     51406124266249856 &        60.116179 &        20.533114 & M5.75 &        1.73 & 3 & D\\
     50447448908539264 &        60.129462 &        19.589142 & F9 &        1.08 & 3 & D\\
    162489749661937792 &        60.808145 &        25.883079 & K4.5 &       0.00 & 4 & D\\
\hline\hline
\end{tabular}
\begin{tablenotes}
\scriptsize{
\item[a] References for the sources. (1) Esplin \& Luhman (2019) \cite{ref3}; \\
(2) Galli et al. (2019) \cite{ref65};(3) Kraus et al. (2017) \cite{ref18};\\
(4) Krolikowski, Kraus \& Rizzuto (2021) \cite{ref37}; (5) This work; \\
(6) Liu et al. (2021) \cite{ref2}; (7) Luhman (2018) \cite{ref16}; \\
(8) Luhman (2023) \cite{ref17}; (9) Zhang et al. (2023) \cite{ref66}.
\item[b] ``G'' marks GYSOs while ``D'' for DYSOs.}
\end{tablenotes}
This table is available in its entirety in machine-readable form.
\end{threeparttable}

\end{table}

\begin{figure}[H]
\centering
\hspace{-0.5cm}
\vspace{-0.2cm}
\setlength{\abovecaptionskip}{0.3cm}
\includegraphics[scale=0.28]{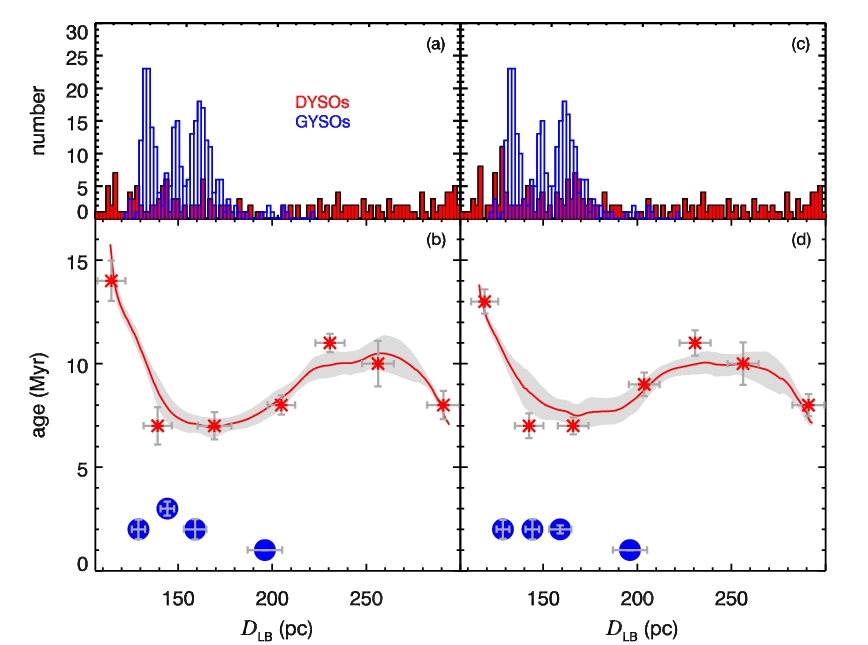}
\caption{Panel (a): The red histogram shows the distance distribution of DYSOs, while the blue histogram indicates the distribution of GYSOs. Panel (b): The red solid line denotes the average age of the DYSOs measured using different sliding window sizes, while the shaded region represents the 1-sigma dispersion. The blue solid dots, along with the error bars, indicate the ages of the GYSOs in the four distance bins and their 1-sigma dispersion from the Monte Carlo simulations, respectively. Panel (c) is the same as Panel (a), and Panel (d) is the same as Panel (b), but for the sample selected using the 20\,Myr isochrone (see the purple solid line in Figure \ref{cds}) instead of the ZAMS curve.}
\label{agedist}
\end{figure}

Due to the lack of radial velocity information, particularly for the newly identified YSOs in this work, it is impractical to conduct a 3D kinematic analysis of the YSOs. On the other hand, since the 2D kinematic analysis of YSOs based on their proper motions has already been thoroughly discussed in the literature \cite{ref2,ref16,ref17,ref37,ref59}, we do not intend to further explore the kinematic information of the YSOs in this region. Instead, to make full use of the sample while ensuring comprehensive sky coverage, we will analyze the YSOs in terms of their distance and age. For this purpose, our classification of the YSO bins will rely solely on their distances to the Sun, which is the center of the Local Bubble. We visually examined the distance distribution of the YSOs. The GYSOs are divided into four bins as they are concentrated within four distinct distance intervals: 125-135\,pc (HD28354, L1524/L1529/B215, L1495/B209), 135-150\,pc (L1527, L1551, T Tau, L1489/L1498), 150-175\,pc (L1517, L1521/B213, L1536, B209N, L1544), and 190-206\,pc (L1558), which can be inferred from Figure \ref{agedist} (see the blue histogram in the upper-left panel). The ages for GYSOs in each bin are derived by minimizing $\chi^{2}$ to the isochrones of PARSEC. Since each bin consists of several ``kinematic groups'', we applied a 100-fold Monte Carlo test to mitigate the effects of member selection. In each run, 80\% of the members of each bin are randomly selected, and their ages are estimated accordingly. We present the results in Figure \ref{agedist} (see the blue solid dots in the lower-left panel). Each data point represents the median of the age fitting and the median distance of the bin members, with the error bars indicating the 1-sigma dispersion.

Unlike GYSOs, DYSOs (including the newly identified YSOs) are sparsely distributed in distance space (see the red histogram in the upper-left panel of Figure \ref{agedist}). Due to the lack of obvious clustering among DYSOs in distance space, the DYSOs are binned into eight equal intervals according to their distances from the Local Bubble. Using the same method applied to the GYSOs, the ages of DYSOs within the distance bins and their 1-sigma deviations are derived and shown as red asterisks and gray error bars in the lower-left panel of Figure \ref{agedist}, respectively. However, we must note that using a single bin size for division is somewhat subjective. Therefore, we also utilized the sliding window method to estimate the ages of DYSOs. We set the window size and treated all DYSOs within the window as a single group. Then, we estimated their ages by minimizing the difference between their locations on the color-magnitude diagram and the isochrones from the PARSEC model. By sliding the window, we obtained the ages of DYSOs relative to their proximity to the Local Bubble. Considering that a single window size is not suitable, we calculated the ages of DYSOs at different positions using various window sizes, ranging from 10 to 40\,pc. In the lower-left panel of Figure \ref{agedist}, we present the average age measurement results with a red solid line, and the shaded area indicates the 1-sigma dispersion of the ages.

We also attempted to replace the zero-age main sequence with an isochrone of 20\,Myr when excluding edge-on disks and variables (see the purple solid lines in Figure \ref{cds}), and the measurement results were essentially consistent (see the right two panels of Figure \ref{agedist}).

\subsection{The formation of the YSOs with respect to the Per-Tau shell and the Tau ring}

\begin{figure}[H]
\centering
\hspace{-0.3cm}
\setlength{\abovecaptionskip}{0.1cm}
\includegraphics[scale=0.23]{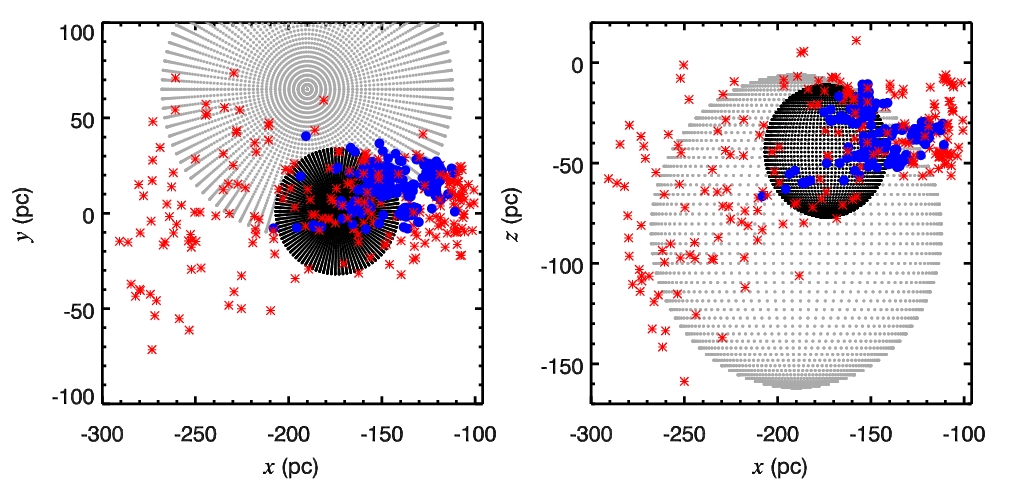}
\caption{The distribution of GYSOs (blue solid dots) and DYSOs (red asterisks) is shown in the $xy$ and $xz$ planes. Also presented are the general shapes of the Taurus ring and the Perseus-Taurus shell, which are denoted by black and gray spheres, respectively.}
\label{xyz}
\end{figure}

\begin{figure}[H]
\centering
\hspace{-0.2cm}
\vspace{-0.1cm}
\setlength{\abovecaptionskip}{0.1cm}
\includegraphics[scale=0.26]{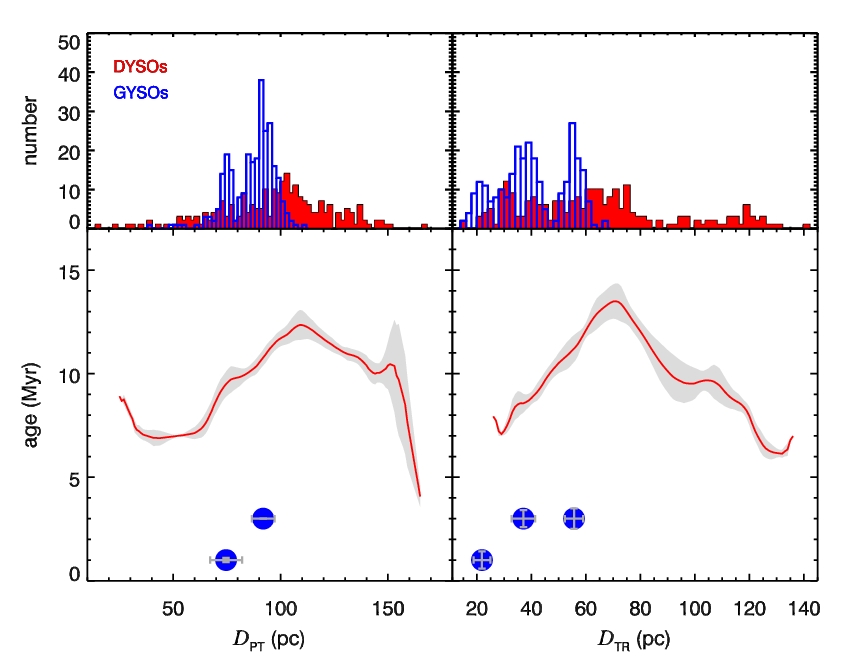}
\caption{Same as Figure \ref{agedist}, but for the Per Tau shell (the left panels) and the Tau ring (right panels).}
\label{PTR}
\end{figure}

Recently, using the 3D dust map of the solar neighborhood \cite{ref61}, Bialy et al. (2021) \cite{ref62} studied the dust density within the extensive Perseus-Taurus region and identified two distinct structures: the ``Per-Tau shell'' and the ``Tau ring''. The Per-Tau shell is a nearly spherical shell located at approximately $(161.1^{\circ}, -22.7^{\circ}, 218\,\text{pc})$ in the Galaxy, with a diameter of approximately $156\,\text{pc}$. Conversely, the Tau ring is elliptical, measuring about $39\,\text{pc}$ in semi-major axis and $26\,\text{pc}$ in semi-minor axis, and is located at approximately $(179.5^{\circ}, -14.2^{\circ}, 179\,\text{pc})$. The authors discussed potential formation scenarios, suggesting that these structures may have arisen from previous stellar and supernova feedback events. Figure \ref{loc} illustrates the Per-Tau shell with a purple dotted line and the Tau ring with a cyan dashed line. YSOs appear to be distributed along the edges of both structures, although this distribution could be a result of projection effects.

To investigate this possibility, a comparison of the spatial locations of YSOs, the Per-Tau shell, and the Tau ring in Cartesian coordinates is presented in Figure \ref{xyz}. The findings suggest that most YSOs are located either inside or outside the Tau ring, rather than along its periphery, indicating that their formation may not be linked to the Tau ring. In contrast, some YSOs seem to follow the periphery of the Per-Tau shell. To verify the relationship between the star formation of Taurus direction and the two structures, we employed the method described in Section \ref{TSD} to measure the relationship between the ages of YSOs and their distances to the Per Tau shell and Tau ring. The result is shown in Figure \ref{PTR}.


\begin{figure}[H]
\centering
\hspace{-0.3cm}
\vspace{-0.1cm}
\setlength{\abovecaptionskip}{0.1cm}
\includegraphics[scale=0.25]{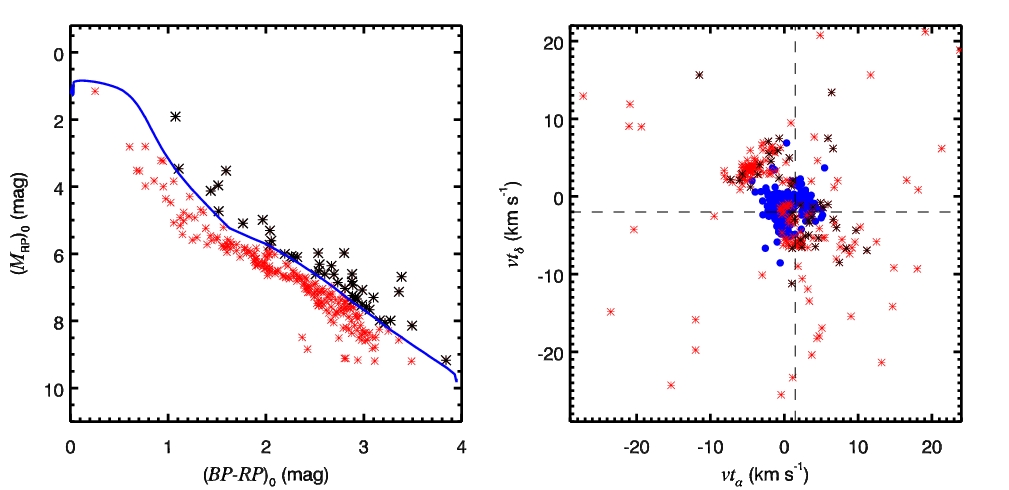}
\caption{The color-magnitude diagram of the DYSOs (red asterisks) is shown in the left panel. The ``CMD-younger'' DYSOs, selected with the 5\,Myr isochrone from PARSEC (blue solid line), are denoted by black asterisks. The right panel depicts the tangential velocities of the YSOs in the Taurus direction. The GYSOs and DYSOs are marked by blue dots and red asterisks, respectively, while the ``CMD-younger'' DYSOs are denoted by black asterisks.}
\label{YDYSOs}
\end{figure}

\section{Discussion}

In Section \ref{TSD}, we explored the relationship between the age and location of YSOs in the Taurus direction and presented the results in Figure \ref{agedist}. The DYSOs generally show a tendency that the farther they are from the Local Bubble, the younger they are, which is consistent with the supernovae-driven formation scenario of the Local Bubble. It is proposed that the DYSOs might be contemporary products of the Local Bubble. In contrast to the DYSOs, GYSOs represent a much younger population, with ages ranging from approximately 1 to 2\,Myr, and their locations are mostly confined to distances between 120 and 220\,pc from the Local Bubble (see the blue solid dots in Figure \ref{agedist}). This places them distinctly apart from the DYSOs in the age-versus-distance space. Therefore, while they may also be associated with the Local Bubble, their formation would appear to be more recent and possibly more localized.

The results also reveal a distinct decrease in the ages of DYSOs between 120 and 220\,pc from the Local Bubble (see Figure \ref{agedist}). Interestingly, this is precisely where the young GYSOs are located. An intuitive explanation is that there is a portion of younger YSOs that coevally formed with the GYSOs, but due to kinematic reasons, they were not recognized and were instead classified as DYSOs. Based on this consideration, we selected stars located above the 5\,Myr isochrone on the CMD (see the left panel of Figure \ref{YDYSOs}) and analyzed their kinematics. In the right panel of Figure \ref{YDYSOs}, we present the tangential velocities of the YSOs, with GYSOs represented by blue solid dots, DYSOs by red asterisks, and the ``CMD-younger'' DYSOs by black asterisks. We can observe that the tangential velocities of GYSOs show a clear clustering trend, mainly distributed around $(-1.5, -2)$\,$\rm km\,s^{-1}$. In contrast, the younger subset of DYSOs, although somewhat closer to the GYSOs, does not exhibit a distinct distribution from the majority of DYSOs. Consequently, based on the available data, we are unable to determine whether these young DYSOs represent scattered GYSOs. Another possible explanation is that these young-looking DYSOs on the CMD are due to photometric variability. Since YSOs are famous for being variable stars, it is reasonable for YSOs to appear in the younger regions on the CMD, but this assumption cannot explain why this phenomenon is particularly significant at distances of 110-220\,pc from the Local Bubble.

We also investigated the correlation between the ages of YSOs and their proximity to the Per~Tau shell and the Tau ring. In Figure \ref{PTR}, the results indicate that there is no strong correlation between the YSOs in the Taurus region and the Per~Tau shell (and the Tau ring), suggesting that these two structures are not decisive factors in the star formation process in the Taurus region. Nevertheless, given that the distribution of the GYSOs outlines the periphery of the Per-Tau shell, it remains challenging to assert that the formation of the GYSOs is entirely independent of the Per-Tau shell's influence.

\section{Summary}

In this work, we conducted a comprehensive survey of new Young Stellar Objects (YSOs) in the Taurus direction. Subsequently, using the newly identified YSOs and the previously identified YSOs, we discussed the relationship between star formation in the Taurus direction and the Local Bubble, the Per~Tau shell, and the Tau ring. The key findings are outlined as follows:
\begin{itemize}

\item
We targeted stars in the Taurus region that are located at distances between 100 and 300~pc from the Sun. We selected stars that lie above the 20~Myr isochrone on the color-magnitude diagram (CMD) as candidate objects. After eliminating the previously identified YSOs, we analyzed the spectra of the remaining candidates with spectral types later than K7. This analysis led to the identification of 141 new YSOs exhibiting distinct Li~{\sc i} absorption lines.

\item
Utilizing photometry from 2MASS and ALLWISE, we investigated the color excess of the YSO candidates and identified four new YSOs that harbor disks.

\item
We examined the spatial distribution and kinematics of the new YSOs in relation to previously identified objects. Our findings suggest that the new YSOs are more likely to be diskless Young Stellar Objects (DYSOs) rather than protostellar Young Stellar Objects (GYSOs).

\item
By combining the previously identified YSOs in Taurus, we established a sample of 519 YSOs. Analyzing their ages in relation to their proximity to the Local Bubble, we discovered a strong correlation between the age of the YSOs and their distance from the Local Bubble ($D_{\rm{LB}}$). Specifically, the further the YSOs are from the Local Bubble, the younger they tend to be. This correlation strongly supports the supernova-driven formation scenario of the Local Bubble.

\item
Using the same method, we discussed the relationship between the YSOs and the Per~Tau shell as well as the Tau ring. The results suggest that the formation of the YSOs may not be correlated with these two structures.

\end{itemize}

\Acknowledgements{M. F. acknowledges the support from the National Key R\&D Program of China with grant 2023YFA1608000. This work is supported by the National Natural Science Foundation of China (NSFC) under grants No. 12003045 and No. 12173013, the Science Foundation of Hebei Normal University under grant No. L2022B07, the project of Hebei Provincial Department of Science and Technology under grant number 226Z7604G, and the Hebei NSF under grant numbers A2021205006 and A2023205036.
The Guoshoujing Telescope (the Large Sky Area Multi-Object Fiber Spectroscopic Telescope, LAMOST) is a National Major Scientific Project built by the Chinese Academy of Sciences. Funding for the project has been provided by the National Development and Reform Commission. LAMOST is operated and managed by the National Astronomical Observatories, Chinese Academy of Sciences. This work has made use of data from the European Space Agency (ESA) mission Gaia (https://www.cosmos.esa.int/gaia), processed by the Gaia Data Processing and Analysis Consortium (DPAC, https://www.cosmos.esa.int/web/gaia/dpac/consortium). Funding for the DPAC has been provided
by national institutions, in particular the institutions participating in the Gaia Multilateral Agreement. Substantial data processing in this work was executed through the TOPCAT software (Taylor 2005).}

\InterestConflict{The authors declare that they have no conflict of interest.}

\renewcommand{\thesection}{Appendix}

\end{multicols}
\end{document}